# Nonlinear geometrically exact dynamics of fluid-conveying cantilevered hard magnetic soft pipe with uniform and nonuniform magnetizations

Amir Mehdi Dehrouyeh-Semnani[1]

School of Mechanical Engineering, College of Engineering, University of Tehran, Tehran, Iran

## Abstract

It is generally acknowledged that a hanging cantilevered pipe conveying fluid becomes unstable by flutter-type instability at a critical flow velocity; moreover, the pipe undergoes periodic self-excited oscillations in the post-flutter region. Additionally, the critical flow velocity increases when the magnetized pipe is exposed to an actuating parallel magnetic field. The question arises as to whether the actuating magnetic field leads to lessening the oscillation amplitude of the system in the post-flutter region. To answer the question, the nonlinear responses of a fluid-conveying cantilevered hard magnetic soft pipe with uniform and nonuniform magnetizations under an actuating parallel magnetic field are examined. In the case of the nonuniform magnetization, the mass density and elastic modulus of the pipe in addition to its residual magnetic flux density vary along its length. The mathematical formulation is constructed via a nonlinear geometrically exact model and is solved by employing the Galerkin technique in conjunction with the Runge-Kutta finite difference scheme. The numerical results are then analyzed to reveal the role of magnetization in the magneto-hydro-elastic responses of the system in the absence and presence of magnetic field.

## 1. Introduction

Due to numerous applications of pipes conveying fluid in engineering systems from macro to small scale, theoretical and experimental investigations on dynamic characteristics of them have received notable considerations [1]. The dynamic analyses of cantilevered pipe conveying fluid based on the third-order nonlinear model have been extensively studied [2-11]. The nonlinear geometrically exact model for simulating the cantilevered soft pipe conveying fluid is now at the center of focus due to the limitation of third-order nonlinear model for large-amplitude motion. Chen et al. [12] developed the geometrically exact equation of motion for the large-amplitude of hanging cantilevered soft pipe conveying fluid. The inaccuracy of third-order nonlinear model for the large-amplitude oscillations of pipe in the post-flutter region was shown in this study. Farokhi et al. [13] conducted a comprehensive numerical investigation on the geometrically exact global dynamics of hanging cantilevered soft pipe conveying fluid in absence and presence of a concentrated mass attached at the free end. They discussed on the importance of geometrically exact model for the global dynamics of pipe under large-amplitude motion.

---

[1] Email addresses: a.m.dehrouye@ut.ac.ir, a.m.dehrouye@gmail.com (A.M. Dehrouyeh-Semnani)

Hard magnetic materials belong to a new class of active materials, consist of a soft polymer matrix embedded with hard magnetic particles, or particles of high coercivity ferromagnetic. They are capable of undergoing ultrafast, extremely large, and complex transforming actuating; moreover, they have the ability of untethered control and outstanding programmability. Therefore, they have received remarkable considerations for design of active structures [14-21]. Additionally, the hard magnetic soft beam as a key element in such active structures has been experimentally and theoretically investigated in many recently published researches. Zhao et al. [22] conducted experimental tests on the extremely large deformation of hard magnetic soft beams under uniform perpendicular and antiparallel magnetic field. The experimental results were compared with those archived based on the three-dimensional finite element simulation. Wang et al. [23] established a theoretical framework for the geometrically exact analysis of hard magnetic soft beams under a uniform actuating magnetic field. The trajectories and workspace of hard magnetic soft beam were comprehensively studied in this work. Wu et al. [24] showed the applications of hard magnetic soft beam-type active structures in one- and two-dimensional structures with asymmetric configuration-shifting to biomimetic crawling robots, swimming robots, and metamaterials with tunable characteristics. Wang et al. [25] designed an optimized workspace for the hard magnetic soft beams under applied uniform magnetic field via a nonuniform magnetization in the beam length direction. Dehrouyeh-Semnani [26] proposed new trajectories and workspace for the hard magnetic soft beams under applied uniform magnetic field due to bifurcation. Chen et al. [27] performed a theoretical examination on the mechanical response of hard magnetic soft beams with designed nonuniform residual magnetic flux density under uniform actuating magnetic field. Yan et al. [28] proposed a comprehensive framework to describe the behavior of hard magnetic soft beams under uniform and constant gradient magnetic fields. The results of beam model were compared with those obtained based on the experimental data and three-dimensional finite element simulation. Wang et al. [29] reported a strategy for the design and optimization of hard magnetic soft beams that can have both a large bending angle and sufficient contact force at the target location.

Chen et al. [30] examined the potential of hard magnetic soft materials for the control of large-amplitude motion of hanging cantilevered pipe conveying fluid. The pipe was magnetized uniformly and partially and was under an applied magnetic field with an inclined angle with respect to the initial configuration of pipe. The stability and nonlinear geometrically exact dynamics of system were numerically studied. Owing to the applied inclined magnetic field, the stable position of pipe was no longer its initial configuration. While a hanging fluid-conveying cantilevered hard magnetic soft pipe with uniform magnetization is exposed to an applied magnetic field with no inclination angle with respect to the magnetization direction, the critical flow velocity at which the system loses its stability by a flutter, increases and the initial configuration of system remains its stable position [30]. The question arises as to whether the system with uniform magnetization in the presence of magnetic field is capable of showing a better performance than the corresponding system in the absence of magnetic field when the flow velocity goes beyond the critical flow velocity. Additionally, it is possible to design a pipe with axially functionally graded magnetization to acquire a better performance in the post-flutter region than the pipe with uniform magnetization. The purpose of this investigation is to respond to these questions.

The outline of the paper is as follows: In the second section, the nonlinear geometrically exact formulation of slender, straight, hanging cantilevered hard magnetic soft pipe with axially functionally graded magnetization under a uniform parallel magnetic field is established in aid of the inextensibility condition and Hamilton's principle. In the third section, the nonlinear integro-partial differential governing equation is discretized via the Galerkin technique and the resulting nonlinear ordinary differential equations are solved by using an embedded Runge-Kutta's formula for the nonlinear dynamic analysis. Additionally, the stability characteristics of system are analyzed based on the linearized governing equation. In the fourth section, the nonlinear geometrically exact responses of a hanging cantilevered pipe conveying fluid and of a hanging fluid-conveying cantilevered hard magnetic soft pipe with uniform magnetization under an applied uniform magnetic field with an inclined angle are used to verify the solution procedure in this paper. In the fifth section, the role of uniform and nonuniform magnetizations in the stability, nonlinear dynamics, and stress distribution of system in the absence and presence of applied magnetic field is examined in detail. The design of magnetic polarity pattern for the nonuniform magnetization is also discussed in this section. The paper is concluded with the sixth section, where the final remarks are provided topic research for future research is suggested.

## 2. Mathematical formulation

The, shown in Figure 1, consists of a slender, straight, hanging cantilevered hard magnetic soft pipe of length $L$, cross sectional area $A$, second moment of area $I$, Young's modulus $E$, mass per unit length m, and residual magnetic flux density $B_0^r$ aligned with the $X$-axis, conveying a fluid of mass per unit length $M$ with mean axial velocity $\overline{U}$, and subjected to an applied uniform parallel magnetic field of magnetic flux density $B^a$. The Young's modulus $E$ and mass density $\rho$ of pipe can be obtained by [25, 31]:

$$E = E_0 \exp\left(\frac{2.5\varphi}{1-1.35\varphi}\right), \tag{1}$$

$$\rho = \rho_m(1-\varphi) + \rho_e\varphi, \tag{2}$$

where $E_0, \rho_m,$ and $\rho_e$ stand for the Young's modulus of a pure polymer without magnetic particles, the mass density of the polymer matrix, and the mass density of the embedded hard magnetic particles, respectively. Additionally, $\varphi$ denotes the magnetic particle volume fraction within the range of $0 \leq \varphi \leq 0.4$ [25]. Without loss of generality, $\varphi$ is defined as follows:

$$\varphi = \varphi_0 \lambda(s^*), \tag{3}$$

in which $\varphi_0$ is $\varphi(s^* = 0)$ and $\lambda$ is the magnetic polarity pattern within the range of $0 \leq \lambda \leq 0.4/\varphi_0$.

In order to develop the mathematical formulation of system, the following assumptions are used: (1) only the planner motion of pipe is considered; (2) the viscoelastic-dynamical model of pipe is derived under isothermal conditions; (3) the pip centerline is supposed to be inextensible; (4) a nonlinear Euler-Bernoulli beam is used to model the pipe, i.e. shear deformation and rotary inertia are ignored; (5) the strain in the pipe is considered small, although large deformations are expected; (6) the constituent material of pipe is incompressible, and its behavior is similar to an isotropic material; (7) the structural damping of pipe obeys the Kelvin-Voight model and it is regarded as a constant value; (8) the fluid is incompressible but not inviscid; (9) the plug velocity profile of flow is constant; (10) the effect of applied magnetic field on the mass density of fluid and the plug velocity profile of flow is ignorable. The deformed beam centerline, $\mathbf{r}$, can be expressed as

$$\mathbf{r}(s^*,t) = x(s^*,t)\mathbf{i} + y(s^*,t)\mathbf{j} = \left(\int_0^{s^*} \cos(\theta(s'))d\overline{s}\right)\mathbf{i} + \left(\int_0^{s^*} \sin(\theta(s')) ds'\right)\mathbf{j}, \tag{4}$$

in which the curvilinear coordinate $s^*$ denotes the distance of a point of the centerline from the origin of the $XY$ coordinate system. Additionally, the xy coordinate system is associated with the $XY$ coordinate system via: $x = X + u$ and $y = Y + v$ in which $u$ and $v$ stand for the longitudinal displacement and transverse displacement, respectively. Furthermore, $\theta(s^*)$ is the angle between the tangent vector of the centerline and the X-axis which is called rotation angle.

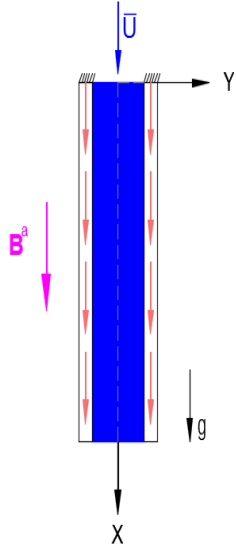

Figure 1: Schematic representation of a hanging cantilevered hard magnetic soft pipe conveying fluid under a uniform parallel magnetic field.

According to the Euler-Bernoulli beam hypothesis and the third assumption, the only non-zero component of strain tensor, $\varepsilon_{xx}$, can be written as

$$\varepsilon_{xx}\left(s^{*},z,t\right)=-z\frac{\partial\theta}{\partial s^{*}} \tag{5}$$

In view of the above equation and the sixth assumption, the only component of stress tensor due to the elastic deformation, $\sigma_{xx}^{e}$, can be expressed as

$$\sigma_{xx}^{e}\left(s^{*},z,t\right)=-zE\left(s^{*}\right)\frac{\partial\theta}{\partial s^{*}} \tag{6}$$

Therefore, the potential energy of system due to the elastic deformation in the pipe, $U_P$, can be obtained by

$$U_{p}=\int_{0}^{L}E\left(s^{*}\right)I\left(\frac{\partial\theta}{\partial s^{*}}\right)^{2}ds^{*}. \tag{7}$$

The potential energy of system due to the gravity, $U_g$, may be expressed as

$$U_{g}=\int_{0}^{L}\left(M+m\left(s^{*}\right)\right)g\left(\int_{0}^{s^{*}}\cos\left(\theta(s')\right)ds'\right)ds^{*}. \tag{8}$$

The kinetic energy of system, $T$, may be written as

$$T=\frac{1}{2}\int_{0}^{L}\left(m\left(s^{*}\right)V_{p}^{2}+MV_{f}^{2}\right)ds^{*}, \tag{9}$$

in which $V_p$ and $V_f$ are the pipe and fluid velocities, defined, respectively, by

$$\mathbf{V}_{p}=\frac{\partial\mathbf{r}}{\partial t}=\dot{x}\mathbf{i}+\dot{y}\mathbf{j}=\left(-\int_{0}^{s^{*}}\frac{\partial\theta}{\partial t}\sin(\theta)ds'\right)\mathbf{i}+\left(\int_{0}^{s^{*}}\frac{\partial\theta}{\partial t}\cos(\theta)\,ds'\right)\mathbf{j},$$

$$\mathbf{V}_{f}=\left(\frac{\partial}{\partial t}+\bar{U}\frac{\partial}{\partial s^{*}}\right)\mathbf{r}=\left(-\int_{0}^{s^{*}}\frac{\partial\theta}{\partial t}\sin(\theta)ds'+\bar{U}\cos(\theta)\right)\mathbf{i}+\left(\int_{0}^{s^{*}}\frac{\partial\theta}{\partial t}\cos(\theta)\,ds'+\bar{U}\sin(\theta)\right)\mathbf{j}. \tag{10}$$

Substituting Eq. (10) into Eq. (9) gives

$$T=\frac{1}{2}\int_{0}^{L}\left(M+m\left(s^{*}\right)\right)\left(\left(\int_{0}^{s^{*}}\frac{\partial\theta}{\partial t}\cos(\theta)ds'\right)^{2}+\left(\int_{0}^{s^{*}}\frac{\partial\theta}{\partial t}\sin(\theta)ds'\right)^{2}\right)ds^{*}$$

$$+\frac{1}{2}M\int_{0}^{L}\left(\bar{U}^{2}+2\bar{U}\cos(\theta)\int_{0}^{s^{*}}\frac{\partial\theta}{\partial t}\cos(\theta)ds'+2\bar{U}\sin(\theta)\int_{0}^{s^{*}}\frac{\partial\theta}{\partial t}\sin(\theta)ds'\right)ds^{*}. \tag{11}$$

In view of the seventh assumption, the virtual work due to the structural damping can be formulated as

$$\delta W_{d}=-\alpha^{*}I\int_{0}^{L}\left(\frac{\partial^{2}\theta}{\partial s^{*}\partial t}\delta\frac{\partial\theta}{\partial s^{*}}\right)ds^{*},$$

in which $\alpha^*$ is the damping coefficient of pipe.

The virtual work due to fluid exiting the pipe, $\delta W_f$, may be written as

$$\delta W_f = \int_{t_1}^{t_2} \left[ M\bar{U} \left( \frac{\partial \mathbf{r}_L}{\partial t} + U\boldsymbol{\tau}_L \right) \delta \mathbf{r}_L \right] dt =$$

$$-\int_{t_1}^{t_2} M\bar{U} \left[ \left( \frac{\partial x(L,t)}{\partial t} + U \frac{\partial x}{\partial s}(L,t) \right) \delta x(L,t) + \left( \frac{\partial y(L,t)}{\partial t} + U \frac{\partial y}{\partial s}(L,t) \right) \delta y(L,t) \right] dt$$

$$-\int_{t_1}^{t_2} M\bar{U} \left[ \left( -\int_0^L \frac{\partial \theta}{\partial t} \sin(\theta) ds' + U \cos(\theta(L)) \right) \left( -\int_0^L \sin(\theta) \delta\theta ds \right) \right.$$

$$\left. + \left( \int_0^L \frac{\partial \theta}{\partial t} \cos(\theta) ds' + U \sin(\theta(L)) \right) \int_0^L \cos(\theta) \delta\theta ds \right] dt, \quad (12)$$

in which $\mathbf{r}_L$ and $\boldsymbol{\tau}_L$ stand for, respectively, the position and tangential unit vectors at the end of the pipe.

The virtual work due to the applied magnetic field, $\delta U_m$, may be expressed as

$$\delta U_m = -\frac{A |B^a| |B_0^r(\varphi_0)|}{\mu_0} \int_0^L \lambda(s^*) \sin(\theta) \delta\theta ds^*, \quad (13)$$

in which $\mu_0$ denotes the vacuum permeability.

The governing equation of system can be obtained by employing the extended Hamilton's principle.

$$\delta \int_{t_1}^{t_2} (T - U_p - U_g) dt + \int_{t_1}^{t_2} (\delta W_f + \delta W_d + \delta U_m) dt = 0. \quad (14)$$

Substituting Eqs. (7), (8), (11), (12), and (13) into Eq. (14) gives

$$-\left( E(s^*) \frac{\partial^2 \theta}{\partial (s^*)^2} + \frac{\partial E(s^*)}{\partial s^*} \frac{\partial \theta}{\partial s^*} \right) I + \sin(\theta) \left( M(L - s^*) - \int_L^{s^*} m(\bar{s}) d\bar{s} \right) g$$

$$+ \cos(\theta) \int_L^{s^*} \left[ (M + m(\bar{s})) \int_0^{\bar{s}} \left( \left( \frac{\partial \theta}{\partial t} \right)^2 \sin(\theta) - \frac{\partial^2 \theta}{\partial t^2} \cos(\theta) \right) ds' \right] d\bar{s}$$

$$- \sin(\theta) \int_L^{s^*} \left[ (M + m(\bar{s})) \int_0^{\bar{s}} \left( \left( \frac{\partial \theta}{\partial t} \right)^2 \cos(\theta) + \frac{\partial^2 \theta}{\partial t^2} \sin(\theta) \right) ds' \right] d\bar{s} \quad (15)$$

$$- 2M\bar{U} \left( \sin(\theta) \int_1^{s^*} \left( \frac{\partial \theta}{\partial t} \right) \sin(\theta) ds' + \cos(\theta) \int_1^{s^*} \left( \frac{\partial \theta}{\partial t} \right) \cos(\theta) ds' \right)$$

$$- \alpha^* \frac{\partial^3 \theta}{\partial (s^*)^2 \partial t} + M\bar{U}^2 \sin(\theta(L) - \theta) + A \mu_0^{-1} |B^a| |B_0^r(\varphi_0)| \lambda(s^*) \sin(\theta) = 0.$$

Introducing the following dimensionless parameters:

$$s = \frac{s^*}{L}, \; \zeta = \frac{u}{L}, \eta = \frac{v}{L}, \; \Im = \frac{E}{E(\varphi_0)}, \; (\tau,\alpha) = \sqrt{\frac{E(\varphi_0)I}{(M+m(\varphi_0))L^4}}(t,\alpha^*), \; U = \left(\frac{M}{E(\varphi_0)I}\right)^{1/2} \bar{U},$$

$$\beta = \frac{M}{M+m(\varphi_0)}, \; \Gamma = \frac{m}{m(\varphi_0)}, \; \gamma = \frac{M+m(\varphi_0)}{E(\varphi_0)I}gL^3, \; \mathrm{P} = \frac{AL^2 |B_0^r(\varphi_0)||B^a|}{E(\varphi_0)I\mu_0},$$

(16)

substituting them into Eq. (15) gives the governing equation in dimensionless form as follows:

$$-\Im(s)\frac{\partial^2 \theta}{\partial s^2} - \frac{\partial \Im(s)}{\partial s}\frac{\partial \theta}{\partial s} + \left(\beta(1-s) - (1-\beta)\int_1^s \Gamma(\bar{s})d\bar{s}\right)\gamma \sin(\theta)$$

$$+ \cos(\theta)\int_1^s \left[(\beta + (1-\beta)\Gamma(\bar{s}))\int_0^{\bar{s}}\left(\left(\frac{\partial \theta}{\partial \tau}\right)^2 \sin(\theta) - \frac{\partial^2 \theta}{\partial \tau^2}\cos(\theta)\right)ds'\right]d\bar{s}$$

$$- \sin(\theta)\int_1^s \left[(\beta + (1-\beta)\Gamma(\bar{s}))\int_0^{\bar{s}}\left(\left(\frac{\partial \theta}{\partial \tau}\right)^2 \cos(\theta) + \frac{\partial^2 \theta}{\partial \tau^2}\sin(\theta)\right)ds'\right]d\bar{s} \quad (17)$$

$$-2U\sqrt{\beta}\left(\sin(\theta)\int_1^s \frac{\partial \theta}{\partial t}\sin(\theta)ds' + \cos(\theta)\int_1^s \frac{\partial \theta}{\partial \tau}\cos(\theta)ds'\right)$$

$$-\alpha \frac{\partial^3 \theta}{\partial s^2 \partial \tau} + U^2 \sin(\theta(1) - \theta) + P\lambda(s)\sin(\theta) = 0.$$

## 3. Solution method

### 3.1. Dynamics

In order to be able to solve the integro-partial differential governing equation, the Galerkin method is utilized to discretize and decrease it into a set of ordinary differential equations. Therefore, the rotational angle $\theta$ is supposed as the following approximate series expansion:

$$\theta = \sum_{n=1}^{N} \phi_n p_n, \quad (18)$$

where $\phi_n$ and $p_n$ represent the $n$th approximate function and the $n$th general coordinates of the rotational motion, respectively. The first derivatives of linear mode shapes of a cantilevered Euler-Bernoulli beam are taken as the approximate functions.

$$\phi_n = \sin(\lambda_n \zeta) + \sinh(\lambda_n \zeta) + \upsilon_n(\cos(\lambda_n \zeta) - \cosh(\lambda_n \zeta)), \quad (19)$$

in which

$$\upsilon_n = \frac{\cos(\lambda_n) + \cosh(\lambda_n)}{\sin(\lambda_n) + \sinh(\lambda_n)}, \; \lambda_1 = 1.8751, \; \lambda_2 = 4.6941, \; \lambda_3 = 7.8547, \; \lambda_n = \frac{(2n-1)\pi}{2} \text{ with } n \geq 4. \quad (20)$$

In order to apply the Galerkin method, Eq. (18) is inserted into Eq. (17) and the subsequent equations are multiplied by the associated approximate function and are integrated with respect to $s$ from 0 to 1; resulting in the following set of ordinary differential equations:

$$M_{ij}(\{p\})\ddot{p}_j + G_{ij}(\{p\})\dot{p}_{ij} + C_{ij}\dot{p}_{ij} + K_{ij}p_{ij} + M^*_{ijk}(\{p\})\dot{p}_j\dot{p}_k \\ + K^*_i(\{p\}) + F_i(\{p\}) = 0, \; i,j,k=1,2,\ldots,N, \quad (21)$$

in which the coefficients are defined as follows:

$$M_{ij} = \int_0^1 \phi_i \left[ -\cos\left(\sum_{n=1}^N \phi_n p_n\right) \int_1^s (\beta + (1-\beta)\Gamma(\bar{s})) \int_0^{\bar{s}} \phi_j \cos\left(\sum_{n=1}^N \phi_n p_n\right) ds' d\bar{s} \right. \\ \left. - \sin\left(\sum_{n=1}^N \phi_n p_n\right) \int_1^s (\beta + (1-\beta)\Gamma(\bar{s})) \int_0^{\bar{s}} \phi_j \sin\left(\sum_{n=1}^N \phi_n p_n\right) ds' d\bar{s} \right] ds,$$

$$G_{ij} = -2U\sqrt{\beta} \int_0^1 \phi_i \left[ \sin\left(\sum_{n=1}^N \phi_n p_n\right) \int_1^s \phi_j \sin\left(\sum_{n=1}^N \phi_n p_n\right) ds' \right. \\ \left. + \cos\left(\sum_{n=1}^N \phi_n p_n\right) \int_1^s \phi_j \cos\left(\sum_{n=1}^N \phi_n p_n\right) ds' \right] ds, \quad C_{ij} = \alpha \int_0^1 \phi_i \frac{\partial^2 \phi_j}{\partial s^2} ds$$

$$K_{ij} = \int_0^1 \phi_i \left( \Im(s) \frac{\partial^2 \phi_j}{\partial s^2} + \frac{\partial \Im(s)}{\partial s} \frac{\partial \phi_j}{\partial s} \right) ds \quad (22)$$

$$M^*_{ijk} = \int_0^1 \phi_i \left[ \cos\left(\sum_{n=1}^N \phi_n p_n\right) \int_1^s ((1-\beta)\Gamma(s') + \beta) \int_0^{\bar{s}} \phi_j \phi_k \sin\left(\sum_{n=1}^N \phi_n p_n\right) ds' d\bar{s} \right. \\ \left. - \sin\left(\sum_{n=1}^N \phi_n p_n\right) \int_1^s ((1-\beta)\Gamma(s) + \beta) \int_0^{\bar{s}} \phi_j \phi_k \cos\left(\sum_{n=1}^N \phi_n p_n\right) ds' d\bar{s} \right] ds$$

$$K^*_i = \int_0^1 \phi_i \left[ \left( \beta(1-s) - (1-\beta) \int_1^s \Gamma(s) d\bar{s} \right) \gamma \sin\left(\sum_{n=1}^N \phi_n p_n\right) + U^2 \sin\left(\sum_{n=1}^N (\phi_n - \phi_n(1)) p_n\right) \right] ds$$

$$F_i = P \int_0^1 \phi_i \lambda(s) \sin\left(\sum_{n=1}^N \phi_n p_n\right) ds$$

In order to obtain the nonlinear responses of system, Eq. (21) is solved by implementation of an embedded Runge-Kutta's formula RK5 (4) established by Dormand and Prince [32].

### 3.2. Stability

The linear form of governing equation around its initial configuration can be obtained by dropping the nonlinear terms from Eq. (17) as follows:

$$-E(s)\frac{\partial^2 \theta}{\partial s^2} - \frac{\partial E(s)}{\partial s}\frac{\partial \theta}{\partial s} + \left(\beta(1-s) - (1-\beta)\int_1^s \Gamma(\bar{s})d\bar{s}\right)\gamma\theta + U^2(\theta(1) - \theta)$$
$$+P\lambda(s)\theta - \alpha\frac{\partial^3 \theta}{\partial s^2 \partial \tau} - 2U\sqrt{\beta}\int_1^s \frac{\partial \theta}{\partial \tau}d\bar{s} - \int_1^s (\beta + (1-\beta)\Gamma(\bar{s}))\left(\int_0^{\bar{s}} \frac{\partial^2 \theta}{\partial \tau^2}ds'\right)d\bar{s} = 0. \tag{23}$$

Substituting Eq. (18) into Eq. (23) and applying the Galerkin technique give

$$\bar{M}_{ij}\ddot{p}_j + (\bar{G}_{ij} + C_{ij})\dot{p}_j + (K_{ij} + K_{ij}^{\gamma} + K_{ij}^{U} + K_{ij}^{m})p_j = 0, \quad i,j=1,2,...,N, \tag{24}$$

in which $C_{ij}$ and $K_{ij}$ are defined in Eq. (22) and the other coefficients are defined as follows:

$$\bar{M}_{ij} = -\int_0^1 \phi_i \left[\int_1^s (\beta + (1-\beta)\Gamma(\bar{s}))\int_0^{\bar{s}} \phi_j ds'd\bar{s}\right]$$

$$\bar{G}_{ij} = -2U\sqrt{\beta}\int_0^1 \phi_i \int_1^s \phi_j \sin\left(\sum_{n=1}^N \phi_n p_n\right)ds'ds$$

$$K_{ij}^{\gamma} = \gamma\int_0^1 \left(\beta(1-s) - (1-\beta)\left(\int_1^s \Gamma(s)d\bar{s}\right)\right)\phi_i\phi_j ds \tag{25}$$

$$K_{ij}^{U} = U^2\int_0^1 \phi_i(\phi_j(1) - \phi_j)ds, \quad K_{ij}^{m} = P\int_0^1 \lambda(s)\phi_i\phi_j ds.$$

Considering the solution of linear oscillation as follows:

$$\mathbf{p}(\tau) = \boldsymbol{\delta} e^{\Omega\tau}, \tag{26}$$

in which $\Omega$ is a complex value and $\delta$ is the oscillation amplitude.

Substituting Eq. (26) into Eq. (24) gives

$$\left(\Omega^2[\bar{\mathbf{M}}] + \Omega([\bar{\mathbf{G}}] + [\mathbf{C}]) + ([\mathbf{K}] + [\mathbf{K}^{\gamma}] + [K^U] + [K^m])\right)\{\boldsymbol{\delta}\} = 0. \tag{27}$$

The critical flow velocity at which the system losses its stability can be obtained by solving the quadratic eigenvalue problem given in Eq. (27).

## 4. Comparative studies

The geometrically exact nonlinear dynamic responses of a cantilevered soft pipe conveying fluid in the post-flutter region reported by Farokhi et al. [13] are utilized to verify the established solution strategy in the current study. Prior to comparison, it should be pointed out that the approximated functions in the Galerkin technique used by the aforementioned references is similar to the current approximate functions (Eq. (19)). The first plot of Figure 2 shows the dimensionless tip transverse displacement for 2, 3, and 4 modes. It can be seen that the predicted results in this study are in excellent agreement with those reported by the abovementioned reference for all the cases. In the second and third plots of Figure 2, respectively, the tip rotation angle and the dimensionless longitudinal displacement are demonstrated. It can be deduced the results of present study with 4 modes verifies well those obtained by Farokhi et al. [13] based on 10 modes.

If the system is subjected to an applied magnetic field with an inclination angle $\bar{\alpha}$ with respect to $X$-axis, the governing equation of the system can be obtained by replacing $P\lambda\sin(\theta)$ with $P\lambda\sin(\theta - \bar{\alpha})$ in Eq. (17). Hence, the geometrically exact nonlinear response of a cantilevered hard magnetic soft pipe conveying fluid with uniform magnetization ($\varphi = \varphi_0$) under an applied magnetic field with inclination angle $\bar{\alpha} = \pi/3$ presented by Chen et al. [30] is employed for comparative studies. The first two plots of Figure 3 display the time histories of dimensionless transverse and longitudinal displacement at the tip in the one period of steady state oscillation. Additionally, the second two plots of Figure 3 show the corresponding phase plane diagrams, respectively. It can be seen that the present results with four modes are in relatively good agreement with those reported by the above-mentioned reference based on four modes. The minor difference between the results of present study with those reported by the aforementioned reference may arise from the cosine functions used as the approximate functions by Chen et al. [12].

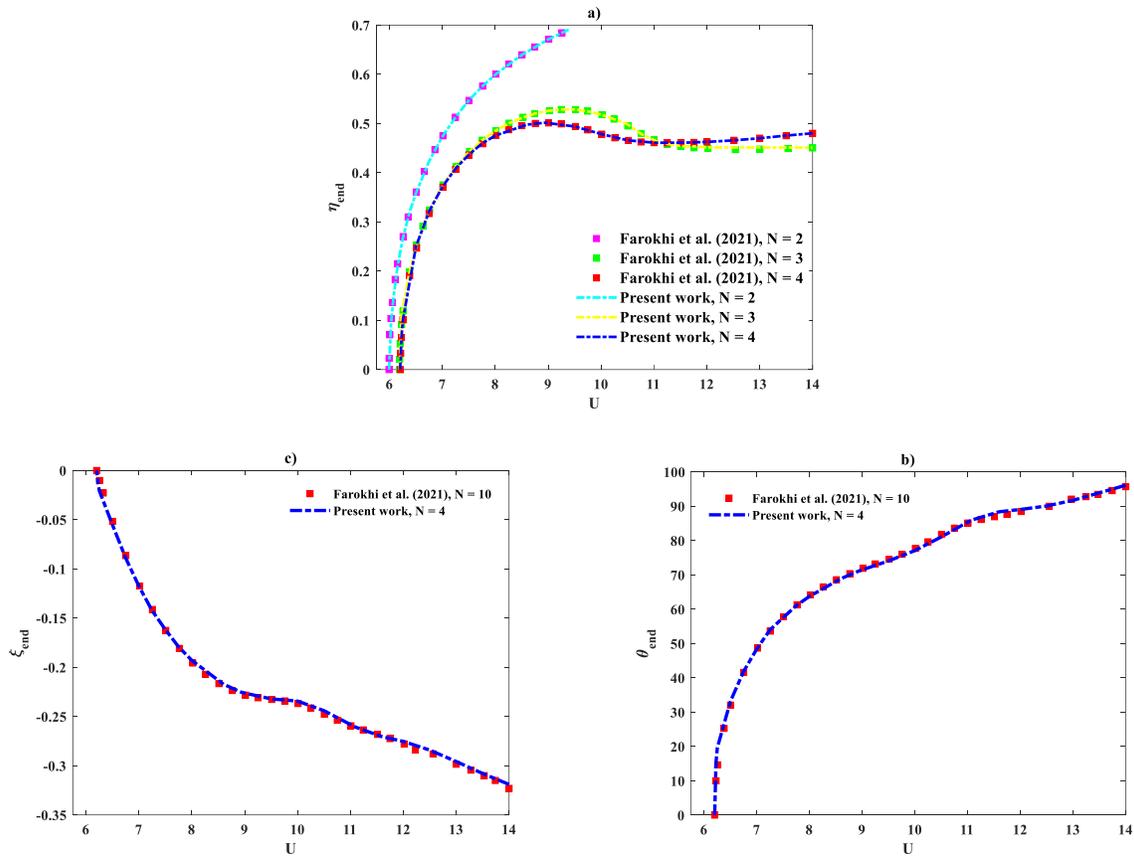

Figure 2: Comparative studies for a soft pipe conveying fluid; a) maximum of dimensionless transverse displacement, b) minimum of dimensionless longitudinal displacement, and c) maximum of rotation angle (in degree) at tip in one period of steady-state oscillation as a function of dimensionless flow velocity when $\beta = 0.142$, $\gamma = 18.9$, and $\alpha = 0.0018$.

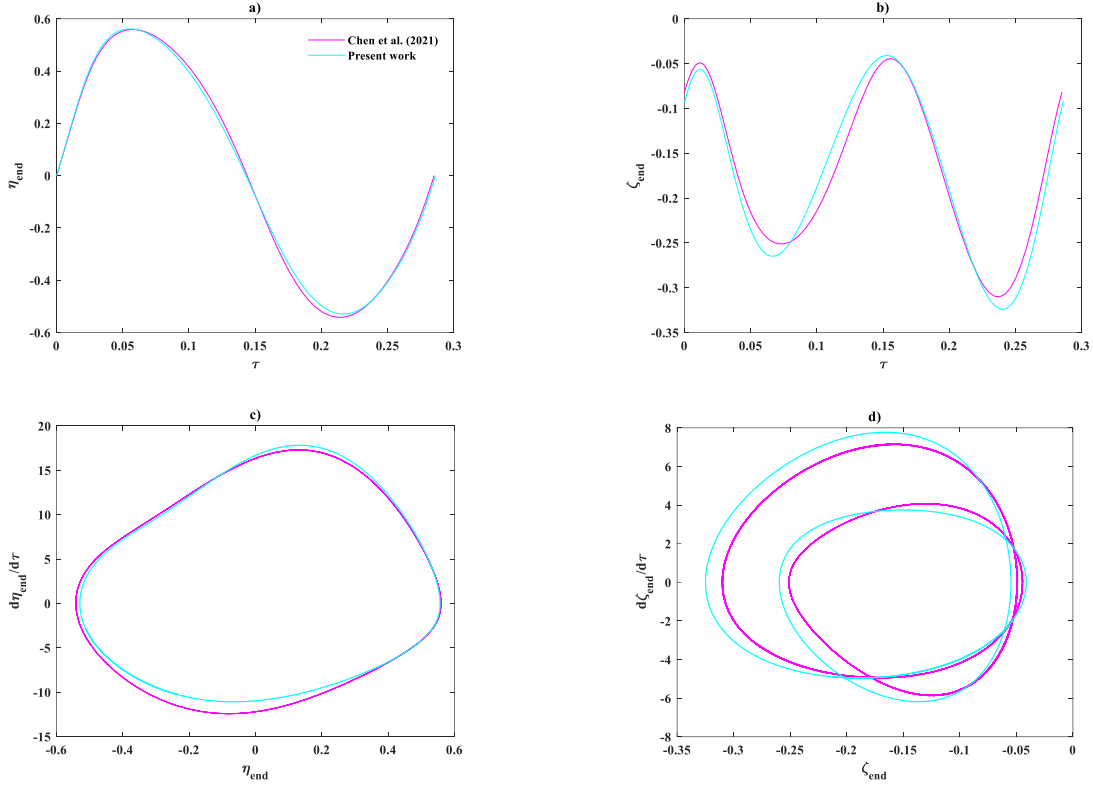

Figure 3: Comparative studies for a fluid-conveying hard magnetic soft pipe with uniform magnetization under an applied magnetic field with inclination angle $\pi/3$ when $\beta = 0.142$, $\gamma = 18.9$, and $\alpha = 0.005$; a) dimensionless transverse and b) longitudinal displacements at tip in one period of steady state oscillation and c, d) corresponding phase plane diagrams, respectively.

## 5. Dynamic analysis

In this section, the influences of an applied uniform parallel magnetic field on the critical flow velocity in addition to the geometrically exact nonlinear dynamics of the hanging cantilevered hard magnetic soft pipe conveying fluid with uniform and nonuniform magnetizations are examined. It should be noticed that in the case of the uniform magnetization, the magnetic particle volume fraction $\varphi = \varphi_0$; therefore, the magnetic polarity pattern $\lambda = 1$. The polymer matrix and the embedded hard magnetic particle are considered to be polyurethane (PDMS) and neodymium-iron-boron (NdFeB), respectively. Therefore, the mass density ratio $\rho_m/\rho_e = 0.1287$. Unless otherwise specified, the following values are considered throughout this study: $\varphi_0 = 0.2$, $\beta = 0.142$, $\gamma = 18.9$; moreover, the dimensionless damping coefficient $\alpha$ is set to 0.0018. In light of the results depicted in the second and third plots of Figure 2, the number of modes in the Galerkin technique is set to 4 for the nonlinear analysis.

Analyzing the longitudinal stress as the only non-zero component of the stress tensor of the pipe was absent in the prior studies related to the nonlinear geometrically exact dynamics of pipe conveying fluid. In order to examine the impact of different parameters on this component of the stress tensor, the dimensionless stress, $\sigma^*$, is defined as follows

$$\sigma^* = -\frac{L}{E(\varphi_0)z}\left(\sigma_{xx}^e + \alpha^* \frac{\partial \varepsilon_{xx}}{\partial t}\right) = \Im\frac{\partial \theta}{\partial s} + \alpha \frac{\partial^2 \theta}{\partial s \partial \tau}, \tag{28}$$

which is related to the dimensionless curvature $\kappa = \partial\theta/\partial s$ and is independent of z coordinate. It should be noticed that the first term of $\sigma^*$ is equal to $\kappa$ for the system with uniform magnetization.

## 5.1. Uniform magnetization

The first plot of Figure 4 shows the effects of the dimensionless magnetic flux density, $P$, within range of $0 \leq P \leq 40$ and the mass parameter, $\beta$, within range of $0 < \beta \leq 0.35$ on the dimensionless critical flow velocity, $U_{cr}$, of the system with uniform magnetization ($\varphi = \varphi_0$). The plotted results indicate that increasing the dimensionless magnetic flux density or the mass parameter leads to a greater dimensionless critical flow velocity. In order to better understand the results of the first plot, the increase rate of dimensionless critical flow velocity, $\delta U_{cr}$, is defined as $\delta U_{cr} = U_{cr}(P,\beta) - U_{cr}(P=0,\beta)$ and is depicted in the second plot of Figure 4.

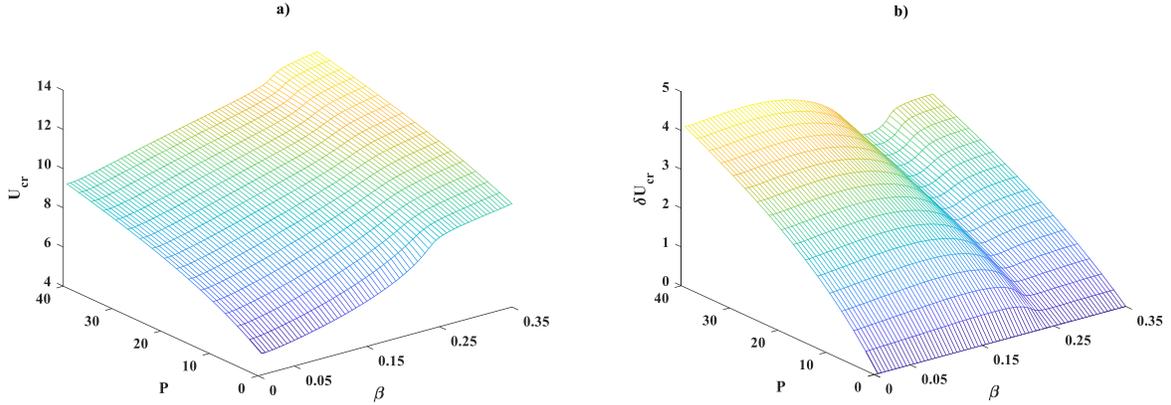

Figure 4: a) Dimensionless critical flow velocity and b) increase rate of dimensionless critical velocity as functions of mass parameter and dimensionless magnetic flux density

As shown in Figure 4, applying a magnetic field leads to increasing the dimensionless critical flow velocity of the system with uniform magnetization. The question arises as to whether the applied magnetic field results in reducing the oscillation amplitude and stress of the system with uniform magnetization in the post-flutter region. To reply the question, the geometrically exact nonlinear dynamics of the system with uniform magnetization in the absence and presence of applied magnetic field is examined in Figure 6. The depicted results with the blue and red colors in Figure 6 show that applying the magnetic field yields a better performance in control of the self-excited oscillation amplitude and stress of the system with uniform magnetization only for a small range of dimensionless flow velocities. It can be concluded that in the post-flutter region, the system with uniform magnetization shows a better performance for a remarkable range of dimensionless flow velocities when the magnetic field is absent. It should be pointed out that this range is dependent on the value of magnetic flux density.

## 5.2. Nonuniform magnetization

In view of Eq. (17), for the system with uniform magnetization, the term due to the applied uniform magnetic field is $Psin(\theta)$ and the term due to the gravity is $(1-s)\gamma\ Psin(\theta)$. Since the gravity results in increasing the critical flow velocity and decreasing the post-flutter oscillation amplitude [13], the magnetic polarity pattern, $\lambda$, is considered to have the following form: $\lambda = 1 - s^n$ in which the degree $n$ is a positive real value. It should be pointed out that in light of Eqs. (1) and (2), and the mass density ratio $\rho_m/\rho_e = 0.1287$, selecting such magnetic polarity pattern leads to reducing both the Young's modulus and mass density of the pipe in the $X$-axis direction. The impacts of degree $n$ on the dimensionless critical flow velocity of the system in the absence and presence of magnetic field are illustrated in the first and second plots of Figure 5, respectively. In view of the plotted results, $n = 5$ is considered for the nonlinear analysis of the system with nonuniform magnetization.

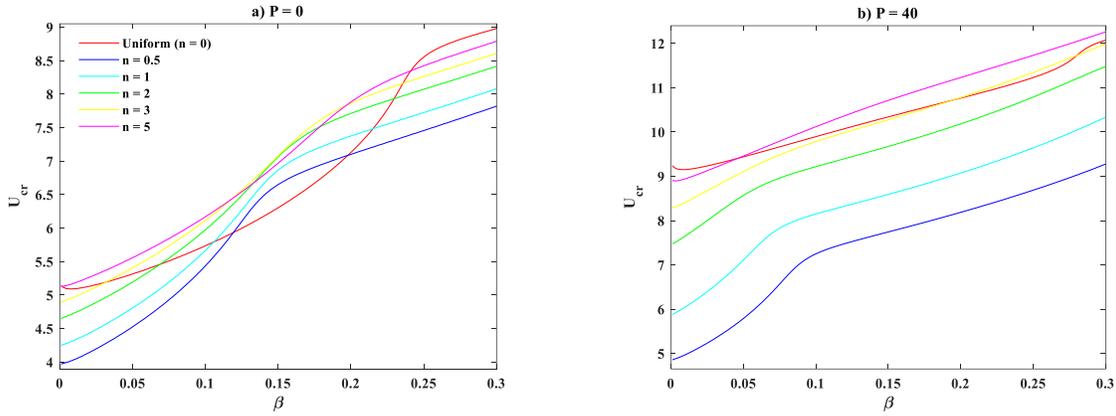

Figure 5: Dimensionless critical flow velocity versus mass parameter for various magnetization patterns in a) absence and b) presence of magnetic field.

The nonlinear steady-state responses of the system with uniform and non-uniform magnetizations in the absence and presence of magnetic field are shown in Figure 6. The obtained results show that the system with non-uniform magnetization is capable of enhancing simultaneously both the critical flow velocity and oscillation amplitude in the presence of magnetic field. Additionally, the system with non-uniform magnetization under applied magnetic field undergoes the least dimensionless stress. While the magnetic field is absent, the tip rotation as well as the tip transverse and longitudinal displacements of the system with non-uniform magnetization may go beyond the corresponding ones of the system with uniform magnetization, but it experiences less dimensionless stress. It's due the fact that the Young's modulus of the system with non-uniform magnetization reduces when $s$ increases.

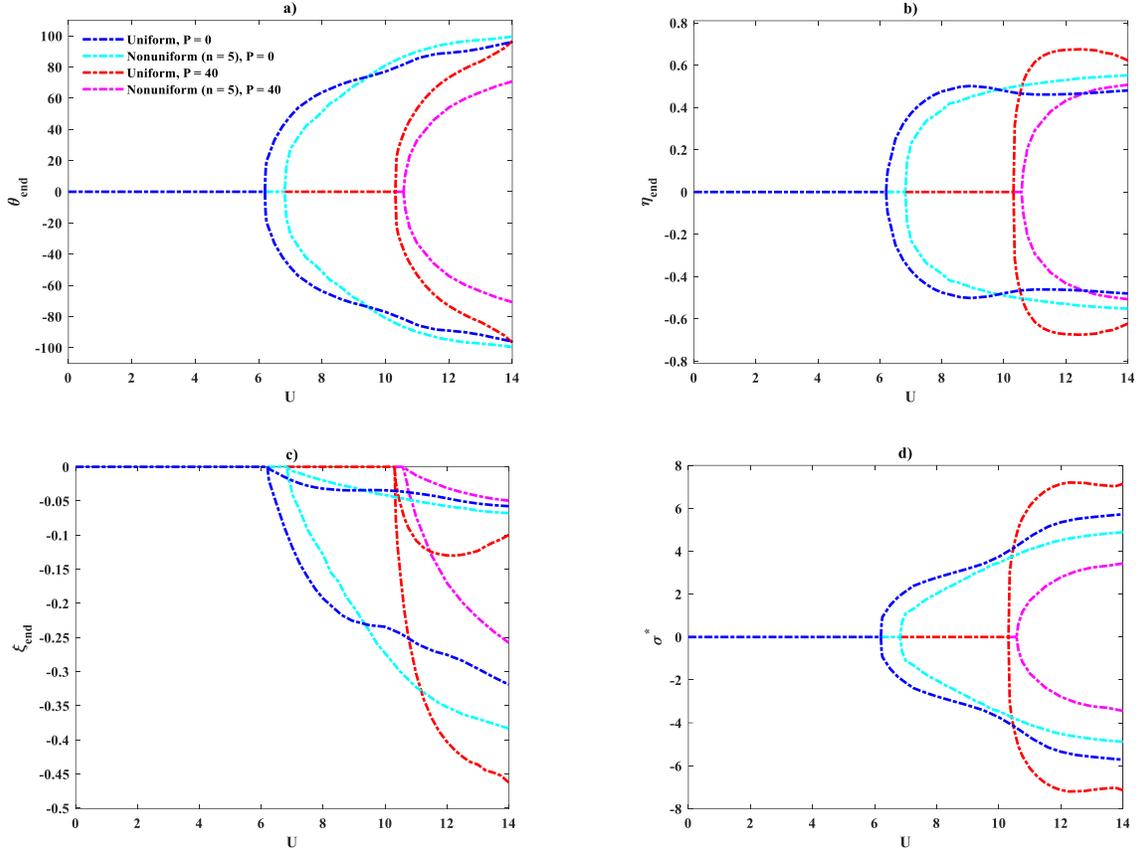

Figure 6: maximum/ minimum values of a) tip rotation angle (in degree), b) dimensionless tip transverse displacement, and c) dimensionless tip longitudinal displacement as well as d) maximum/ minimum values of dimensionless stress in one period of steady-state oscillation for uniform and nonuniform magnetizations and different dimensionless flow velocities in absence and presence of magnetic field.

Figure 7 displays the time histories over a cycle of steady-state oscillation of the tip rotation and corresponding tip transverse and longitudinal displacements of the system with uniform and nonuniform magnetizations when the dimensionless flow velocity $U = 11, 14$ and the dimensionless magnetic flux density $P = 0, 40$. Additionally, the corresponding phase plane diagrams are plotted in Figure 8. The results depicted in Figure 7 illustrate that applying the magnetic field results in increasing the oscillation period and the increase rate for the system with uniform magnetization is remarkably greater than that with nonuniform magnetization. Furthermore, the oscillation periods of the systems with uniform and nonuniform magnetization in the absence of magnetic field are very close. As expected [12], the oscillation periods of the rotation angle and the transverse displacement are the same and are twice of the oscillation period of the longitudinal displacement.

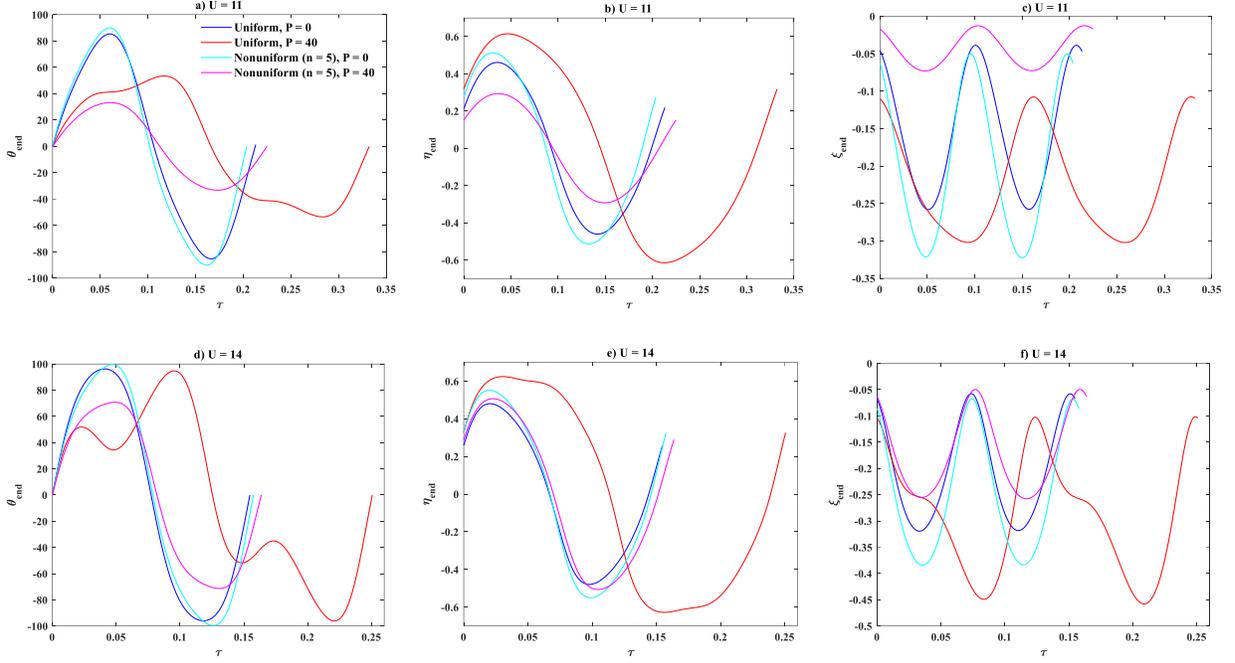

Figure 7: Time traces of tip rotation (in degree) and corresponding tip displacements in one period of steady-state oscillation for uniform and nonuniform magnetizations and different dimensionless flow velocities in absence and presence of magnetic field.

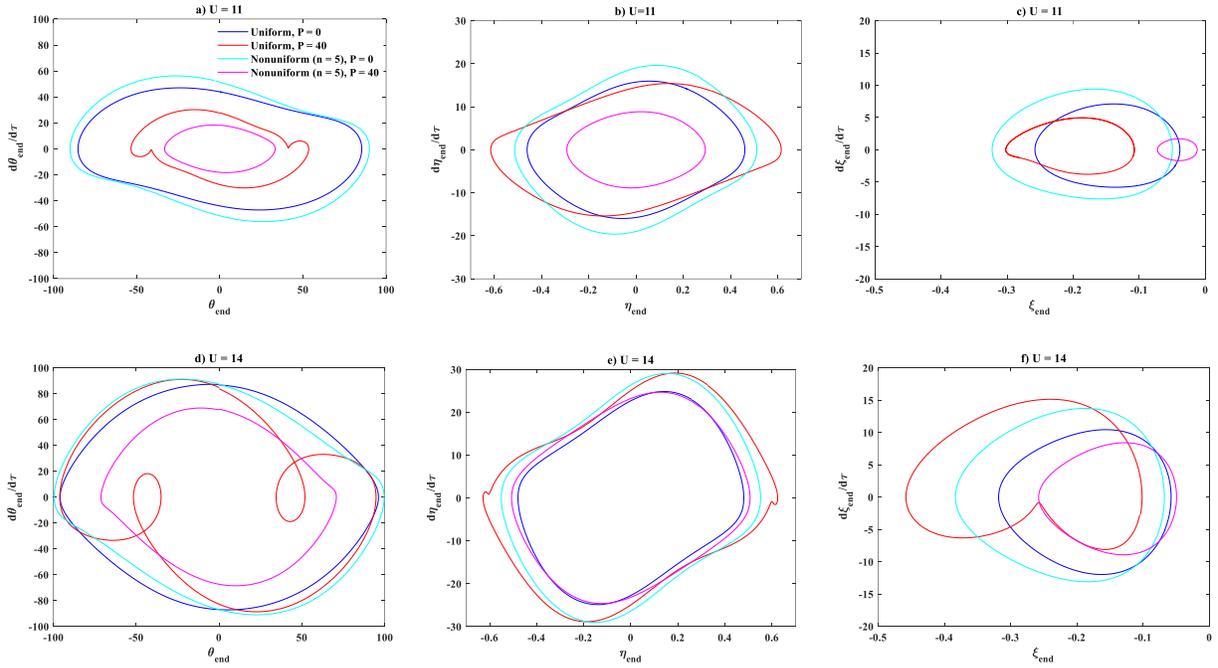

Figure 8: Phase plane diagrams of $\theta$ (in degree), $\eta$, and $\zeta$ at tip for uniform and nonuniform magnetizations and different dimensionless flow velocities in absence and presence of magnetic field corresponding to Figure 7.

In order to better capture the evolution of tip motion, the trajectories of the system with uniform and non-uniform magnetizations in the absence and presence of magnetic field are depicted in Figure 9.

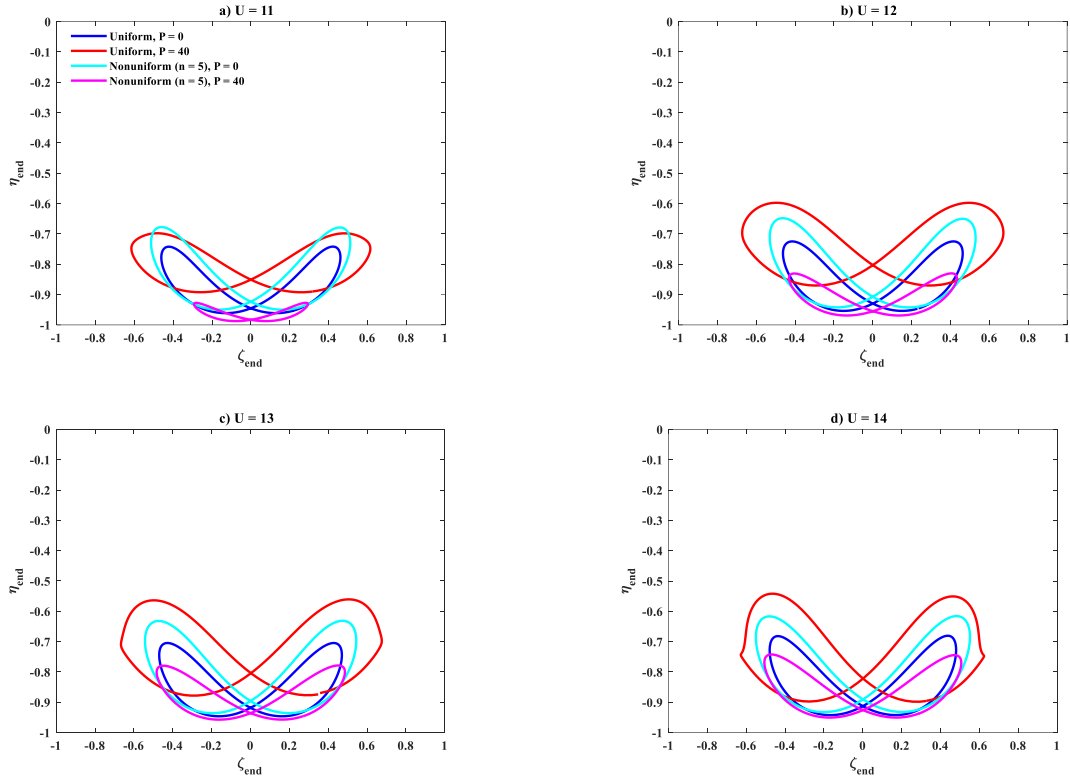

Figure 9: Trajectories of tip over a cycle of steady-state oscillation for uniform and nonuniform magnetizations as well as for different values of dimensionless flow velocities and dimensionless magnetic flux densities.

In order to better understand the pipe motion, for the dimensionless flow velocities $U = 11, 14$ and dimensionless magnetic flux density $P = 0, 40$, the deformed configurations of the system with uniform and non-uniform magnetizations are depicted in Figure 10.

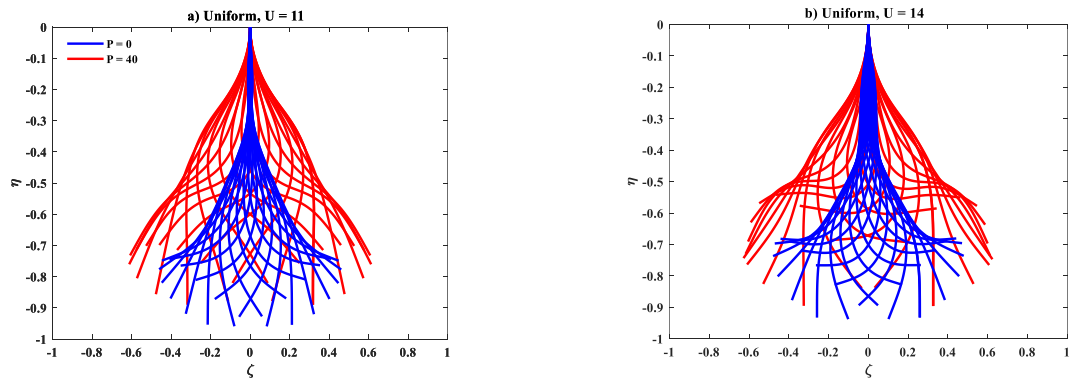

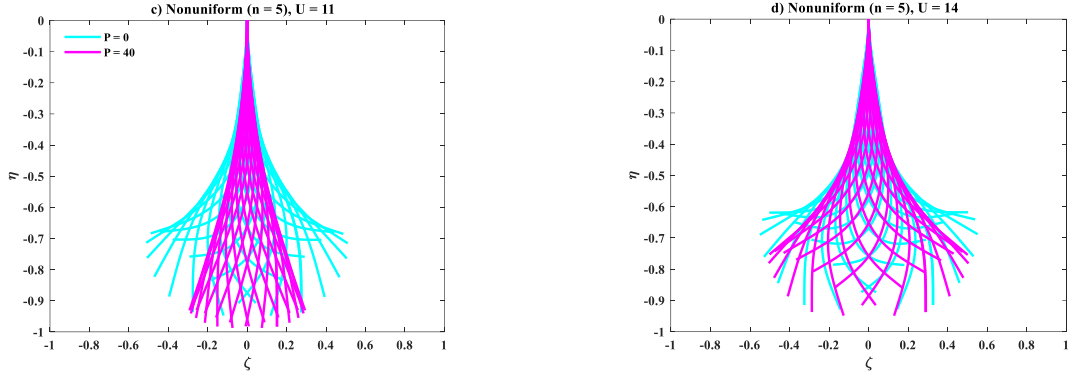

Figure 10: Deformed configurations over a cycle of steady-state oscillation for uniform and nonuniform magnetizations as well as for different values of dimensionless flow velocities and dimensionless magnetic flux densities.

As shown in Figure 11, the dimensionless stress along the pip length in one period of steady-state oscillation for the system with uniform and nonuniform magnetization in the absence and presence of magnetic field when the dimensionless flow velocity $U = 12$. As expected, for all the cases, there is no stress at the pipe tip i.e. $\sigma(1, z, \tau) = 0$. The plots in the figure illustrate that the stress distribution along the pip length is dependent on both the magnetization type and magnetic flux density. Additionally, the time and location at which the maximum or minimum dimensionless stress takes place, is dependent on the two above-mentioned factors.

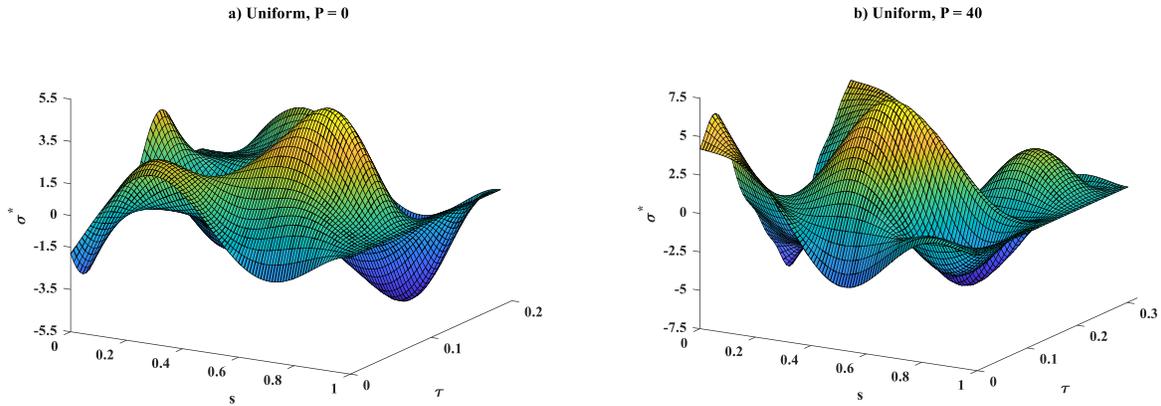

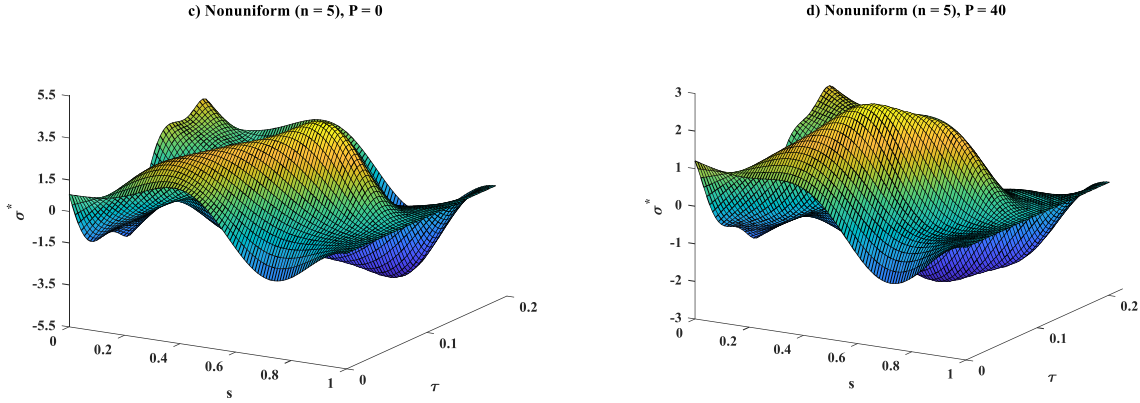

Figure 11: Dimensionless stress over a cycle of steady-state oscillation for uniform and nonuniform magnetizations as well as for different values of dimensionless magnetic flux densities when $U = 12$.

## 6. Conclusion

In the framework of this research, the reliability of fluid-conveying cantilevered hard magnetic soft pipe under an applied parallel magnetic field in the post-flutter region is investigated. The nonlinear geometrically exact mathematical model of system with uniform magnetization as well as with axially functionally graded magnetization was developed and was numerically solved by aid of the Galerkin and Runge-Kutta finite difference methods. The comparative studies indicates the reliability of solution procedure employed in the current study. The case under investigation is considered to be subjected to a high value of the magnetic flux density which is remarkably capable of increasing the critical flow velocity at which the system undergoes a flutter instability. The obtained results show that the system with uniform magnetization in the presence of magnetic field only for a small range of critical flow velocities has a better performance than the corresponding system in the absence of magnetic field. Therefore, in the post-flutter region, the system with uniform magnetization in the presence of magnetic field is unreliable. In analogy of the terms due to the gravity force and applied magnetic field, the nonuniform magnetization is considered to have the following magnetic polarity pattern $\lambda = 1 - s^n$ along the pip length. Based on the analysis of critical flow velocity, the degree $n$ is considered to be 5. It is indicated that the system with axially functionally graded magnetization not only increases remarkably the critical flow velocity but also is capable of lessening both the oscillation amplitude and longitudinal stress in the post-flutter region when the magnetic field is applied. Optimization of the magnetic polarity pattern to obtain a greater critical flow velocity along with smaller oscillation amplitude and longitudinal stress in the post-flutter region is an important topic for future research.